\documentclass[prd,a4paper,nofootinbib,
showpacs, 
]{revtex4}
\usepackage{graphicx}
\usepackage{amsfonts}
\usepackage{amssymb}
\usepackage{slashed} 
\usepackage{color}
\def\eq#1{{eq.~(\ref{#1})}}
\def\eqs#1#2{{eqs.~(\ref{#1})--(\ref{#2})}}

\def\Tr{\mbox{Tr}\,}

\newcommand{\be}{\begin{equation}}
\newcommand{\ee}{\end{equation}}
\newcommand{\bea}{\begin{eqnarray}}
\newcommand{\eea}{\end{eqnarray}}
\newcommand{\nn}{\nonumber}

\begin{document}
%%%%%%%%%%%%%%%%%%%%%%%%%%%%%%%%%%%%%%%%%%%%%%%%%%%%%%%%%%%  FRONT PAGE
\title[]{Stringent limits on top-quark compositeness \\ from $t \bar t$ production at the Tevatron and the LHC}
\date{\today}
\author{M. Fabbrichesi$^{\dag}$}
\author{M. Pinamonti$^{\ddag}$}
\author{A. Tonero$^{\circ}$}
%\author{XYZ}
\affiliation{$^{\dag}$INFN, Sezione di Trieste}
\affiliation{$^{\ddag}$INFN, Sezione di Trieste, Gruppo collegato di Udine and SISSA, via Bonomea 265, 34136 Trieste, Italy}
\affiliation{$^{\circ}$ICTP South American Institute for Fundamental Research \& 
Instituto de F\'{\i}sica Te\'orica UNESP\\
%UNESP - Universidade Estadual Paulista \\
Rua Dr.~Bento T.~Ferraz 271  -  01140-070  S\~ao Paulo, SP, Brazil\\}
%\affiliation{$^{\circ}$ ICTP-SAIFR, Rua Dr.\ Bento Teobaldo Ferraz 271, 01140-070 S\~ao Paulo, Brazil }
\begin{abstract}
\noindent  If the top quark is a composite state made out of some constituents, its interaction with the gluon will be modified. We introduce the leading effective operators that contribute to the radius and anomalous magnetic moment of the top quark and study their effect on the cross section  for $t \bar t$ production  at the Tevatron and  the LHC. Current measurements of the cross sections set a stringent  limit on the scale of compositeness. This  limit is  comparable to   similar limits obtained for light quarks and those from electroweak precision measurements. It  can be used to constrain  the parameter space of some composite Higgs models.
\end{abstract}

%\keywords{}
%%%%%%%%%%%%%%%%%%%%%%%%%%%%%%%%%%%%%%%%%%%%%%%%%%%%%%%%%%%%%%%%%%%
\maketitle
%%%%%%%%%%%%%%%%%%%%%%%%%%%%%%%%%%%%%%%%%%%%%%%%%%%%%%%%%%%%%%%%%%%
%\vskip1.5em
\section{Motivations}
\label{sec:mot}
Whether the top quark is a point-like particle or an extended object is a question that can now  be addressed thanks to the large number of them produced at the LHC and the Tevatron. The most recent, combined measurement of the cross section for $t\bar t $
production is in good agreement with the most up-to-date theoretical prediction within the standard model (SM) and this result can be used to put constraints on the compositeness of the top quark.

Whereas in the SM the top quark---as well as all other fundamental particles---has no structure, various extensions of the  SM are based on some form of compositeness: in particular, the composite Higgs boson~\cite{compo,Agashe:2004rs}, as well as its partial compositeness implementation~\cite{Kaplan:1991dc,Grossman:1999ra}, but also models inspired by the littlest Higgs~\cite{ArkaniHamed:2002qy} and technicolor~\cite{Farhi:1980xs} assume the existence of a strongly interacting sector, the SM particles being either composite object themselves or mixing with particles which are. The top quark, it being the heaviest of all states, is the best candidate for searching for possible signatures of such compositeness. The problem has been  previously addressed in \cite{Georgi:1994ha} and  more recently in \cite{Agashe:2005vg,Giudice:2007fh,Pomarol:2008bh,Lillie:2007hd,kamenik,hioki,Englert:2012by,Degrande:2010kt}. We discuss in some detail the implications of the limits we find on possible extensions to the SM in the last section.

\subsection{Composite top quark and strong interactions}

Compositeness can manifest itself in various ways. We take the most direct approach and look into  what effect a finite extension  of the top quark has on its interaction with  the gauge bosons. In the case most relevant for collider physics, we can write  two form factors $F_1(q^2)$ and $F_2(q^2)$ modifying the  vertex between the top quark and the gluon as
\be
g_s \bar t \left[ \gamma^\mu F_1(q^2)  +  \frac{i \sigma^{\mu\nu}q_\nu}{2 m_t} F_2(q^2) \right]   \, G_\mu \,t \, , \label{int}
\ee
where $g_s$ is the strong $SU(3)_C$ coupling constant, $G_{\mu}= T_A G_{\mu}^A$ is the gluon field, $T_A$ are the $SU(3)_C$ group generators, $q^\mu$ is the momentum carried by the gluon, $t$ denotes the top quark field and $\sigma^{\mu\nu}=i[\gamma^\mu,\gamma^\nu]/2$. The interaction in \eq{int} is the most general after assuming that the vector-like nature of the gluon-top quark vertex  is preserved by the underlying dynamics  giving rise to the composite state. 

As originally pointed out for the case of electromagnetic interactions~\cite{Ernst:1960zza}, the physics of the form factors in \eq{int} is best represented by the combinations
\be\label{gem}
G_E (q^2) = F_1(q^2)+ \frac{q^2}{4 m_t^2} F_2(q^2)  \quad \mbox{and} \quad G_M(q^2) = F_1(q^2) + F_2(q^2) \, ,
\ee
which  are (in the Breit frame) the Fourier transform of, respectively, the chromo-electric and -magnetic charge densities $\rho_c$ and  $\mu \rho_m$ of, in our case, the top quark. For an extended object these densities are not Dirac $\delta$-functions and can be expanded. To the leading order, we thus obtain a first momentum (the chromo-magnetic moment $\mu$):
\be
 G_M( q^2) =  \frac{2}{\pi} \int dr\,  r^2 j_0(qr)  \mu \rho _m(r) \simeq  \mu + \cdots \label{R}
\ee
from the chromo-magnetic charge density, and a second momentum (the squared mean radius $\langle \vec r^2 \rangle$):
\be
G_E( q^2) =  \frac{2}{\pi} \int dr\,  r^2 j_0(qr) \rho _c(r)  \simeq 1 - \frac{\vec q^2}{6} \langle \vec r^2 \rangle + \cdots \, \label {MU}
\ee
from the chromo-electric charge density. In \eqs{R}{MU},  $j_0(x)=\sin x/x$ represents the spherical Bessel function of order zero and the (non-relativistic) charges  $\rho _c$ and $\rho _m$ are related to the four-current as
\be
 j^\mu(r) = \left(g_s \rho_c(r), \, \mu\ \vec \sigma \times \vec \nabla \rho_m(r)\right) \, .
\ee
The two parameters $\mu$ and $\langle \vec r^2 \rangle$ are traditionally used in nuclear physics to characterize the finite extension of nucleons and other extended objects.
 
Form factors are just a way of organizing the perturbative expansion. An alternative and perhaps better approach is effective field theory. In this language the expansion is given in terms of  operators  invariant under  the underlying symmetries  that are added to the SM lagrangian. These operators have dimension higher than four and are suppressed by negative powers of the new physics scale to get the required dimension.

In this work we use the effective field theory approach and consider the contributions given by $SU(3)_C\times U(1)_{em}$ invariant effective operators to the top-quark form factors introduced in \eq{int}. The leading contributions come from the following two higher dimensional operators:
\be\label{O}
\hat O_1 = g_s \frac{C_1}{m_t^2} \bar t \gamma^\mu T_A t D^\nu G^A_{\mu\nu} \quad \mbox{and} \quad \hat O_2= g_s \frac{C_2\upsilon}{2 m_t^2 } \bar t \sigma^{\mu\nu} T_A t G^A _{\mu\nu}
\ee
where $D^\nu G^A_{\mu\nu}=\partial^\nu  G^A_{\mu\nu} + g_s f^A{}_{BC}  G^{\nu B}  G^C_{\mu\nu}$, $G^A_{\mu\nu}=\partial_\mu G_\nu^A-\partial_\nu G_\mu^A +g_s f^A{}_{BC}  G_\mu^B G_\nu^C$ is the gluon field strength tensor, $f^A{}_{BC}$ are the $SU(3)_C$ structure constants and $\upsilon=174$ GeV is the electroweak (EW) symmetry breaking vacuum expectation value. In \eq{O} the operator $\hat O_1$ and $\hat O_2$ are, respectively, of dimension six and five. We limit ourselves to the $CP$-conserving case and the dimensionless coefficients $C_1$ and $C_2$ are taken to be real. Left- and right-handed fermion fields enter symmetrically. The operator $\hat O_1$ gives the leading $q^2$ dependence to $F_1$ while $\hat O_2$ gives the $q^2$-independent term of $F_2$:
\be 
\label{f12}
F_1(q^2)=1+C_1 \frac{q^2}{m_t^2} +\ldots\qquad\mbox{and}\qquad F_2(0)=2\, C_2 \frac{\upsilon}{m_t} \,.
\ee

Operators of higher dimensions can in general contribute---they gives further terms in the expansion of the form factors---but their effect should be  suppressed. We have checked that possible corrections due to dimension 8 operators such like 
\be
G_{\mu\nu} G^{\mu\nu} \bar q \tilde H t_R
\ee
are suppressed as long as the coefficients are taken to be $O(C_{1,2}^2)$.

The form of the coefficients in front of the operators in \eq{O} is conventional and dictated, in our case, by the analogy with the electromagnetic form factors. In addition, the operator $\hat O_2$ is written for convenience with an extra factor $\upsilon/m_t$ because it can be thought as coming, after EW symmetry breaking, from a dimension six $SU(3)_C\times SU(2)_L\times U(1)_Y$ gauge invariant operator that includes the Higgs boson field. 

Replacing the expressions of the form factors obtained in \eq{f12} into \eq{gem}, we can obtain an estimate of the radius and chromo-magnetic moment of the  top quark using the same formulas that apply in the electromagnetic case. We have that
\be\label{rm}
\langle \vec r^2 \rangle = -  6 \left .\frac{d G_E}{d \vec q^2} \right|_{q^2=0} \,\quad \mbox{and} \quad \mu  = G_M(0)\,,
\ee
where $\mu$ is the chromo-magnetic moment of the top quark measured in units of $g_s/2m_t$.

The aim of this work is to give an estimate of the size of these quantities by constraining the values of the dimensionless coefficients $C_1$ and $C_2$ using the available LHC public data on the $t \bar t$ total production cross section $\sigma(pp \rightarrow \bar t t)$ and those for the $\sigma(p\bar p \rightarrow \bar t t)$ from the Tevatron. In addition, we also include constraints from data on spin correlations of the top quarks at the LHC. The results will be used in the  last section where we will  translate the bounds on $C_1$ and $C_2$ into limits on new physics scales in the framework of some specific models.

The finite extension of the source generating the terms in \eq{rm} arises because of the radiative corrections of the parton-level processes as well as because of the presumed compositeness.
In order to disentangle these two contributions we assume that the former is included in the SM cross section computed at the next-to-leading order (NLO) and beyond (for the most recent exact computation at the NNLO and NNLL, see~\cite{Czakon:2013goa} and citations therein), leaving $C_1$ and $C_2$ to encode only  effects intrinsically due to compositeness.

\begin{figure}[ht!]
\begin{center}
\includegraphics[width=4in]{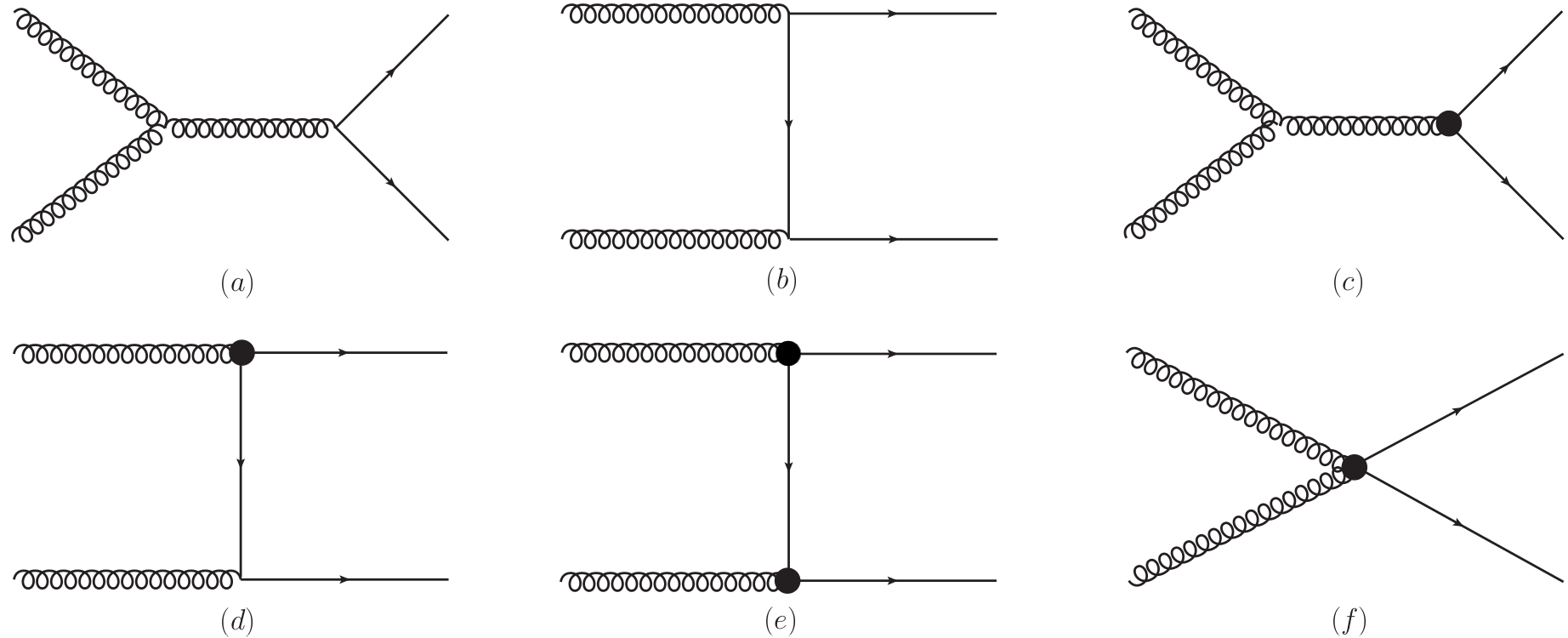}
\caption{\small  Parton level Feynman diagrams for the process $gg\rightarrow \bar tt$. This process dominates at the LHC. The blob represents the insertion of either the operator $\hat O_1$ or $\hat O_2$.
\label{diagrams}}
\vspace{0.5cm}
\includegraphics[width=3in]{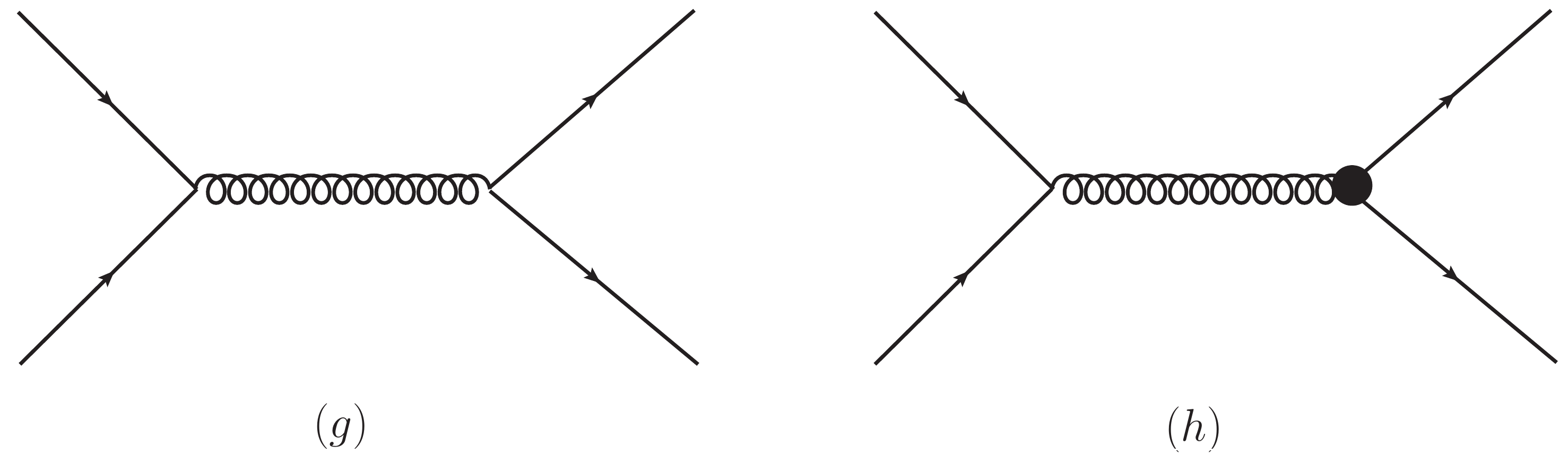}
\caption{\small  Parton level Feynman diagrams for the process $qq\rightarrow \bar t t$. This process dominates at the Tevatron. The blob represents the insertion of either the operator $\hat O_1$ or $\hat O_2$.
\label{diagrams2}}
\end{center}
\end{figure}

The operators $\hat O_1$ and $\hat O_2$ enter at tree level into the computation of the $t \bar t$ production cross section through gluon fusion and quark-antiquark annihilation. The first channel is the dominant one at the LHC while the second is dominant at  the  Tevatron. At tree level, in addition to the usual SM QCD Feynman diagrams, one has to take into account the contributions coming from the new interactions as depicted in Fig.~\ref{diagrams} and Fig.~\ref{diagrams2}. 
The diagram (f) of Fig.~\ref{diagrams} is a contribution not present in the SM and due to the presence of the operators $\hat O_1$ and $\hat O_2$; it represents the effective interaction of two gluons and the $\bar t t$ pair. Notice that the contribution of the operator $\hat O_1$ cancels out in the sum of the gluon fusion amplitudes.

%%%%%%%%%%%%%%%%%%%
%%%%%%%%%%%%%%%%%%%%%%

Following the effective field theory approach, one has to write down all possible dimension five and six operators that contribute to the $t \bar t$ production cross section. It is possible to show~\cite{operators} that---by rearrangement and field redefinitions using the equations of motion---out of all possible operators that contribute to this process only three are independent, namely $\hat O_1$, $\hat O_2$ in \eq{O} and a set of four-fermion operators. If we further assume the same coefficient in front of the four-fermion operators involving two top quarks and two light quarks of different flavors, then these four-fermion operators too can be rewritten, by means of the equations of motion, in terms of the operator $\hat O_1$. We are thus left with just the two operators $\hat O_1$ and $\hat O_2$ in \eq{O}. Notice that in general, the whole set of four fermion operators entering in the dimension 6 SM effective lagrangian is larger and cannot be rewritten as $O_1$.

Anomalous couplings and top quark production has been discussed by several 
authors~\cite{Lillie:2007hd}, most recently in \cite{kamenik} (and \cite{hioki}, which came out while we were finishing this work).   The anomalous magnetic moment of these references corresponds to the coefficient $C_2$ of the chromo-magnetic operator in \eq{O}. 
Ref.~\cite{Englert:2012by} follows an approach similar to ours but with different and less stringent results. The effect on $\bar tt$-production of $\hat O_2$ (together with the four-fermion operators) has been  studied in \cite{Degrande:2010kt} in the context of non-resonant new physics at the Tevatron and the LHC. See \cite{Contino:2013kra} for and updated version.

In the low-energy regime, data on $B$-physics can be affected by the operators in \eq{O}. In particular, the operator $O_2$ contributes to the matching condition of the Wilson coefficient of the chromo-magnetic operator  between the quark $b$ and $s$. The latter operator  mixes with the electro-magnetic dipole moment operator
\be
e\, m_b \,\bar b \, \sigma^{\mu\nu} (1 + \gamma_5 )\, s\, F _{\mu\nu} \, ,
\ee
which give rise to the transition $b\rightarrow s \gamma$.  Even though data on the branching fraction~\cite{cleo} can in principle be used to set   limits on the coefficient $C_2$, the estimates in literature give  either too small an effect~\cite{Hewett:1993em} or one with large uncertainties~\cite{Martinez:2001qs}. For this reason we do not use  these limits.

\section{Methods}

%%%%%%%%%%%%%%%%%%%%%%%%%%%%%%%%%%%%%%

\subsection{Monte Carlo implementation}
\label{sec:mc}
In order to study new physics effects on $t \bar t$ production cross sections (and spin correlations) at the LHC and Tevatron , we have first used {\sc FeynRules}~\cite{Christensen:2008py} to implement our model, which has been defined to be the SM with the addition of the two effective operators $\hat O_1$ and $\hat O_2$ of \eq{O}. {\sc FeynRules} provides the Universal FeynRules Output (UFO) with the Feynman rules of the model. The UFO is then used by {\sc MadGraph 5} \cite{madgraph} (MG5)  to compute the production cross section, that we denote by $\sigma_{\mbox{\scriptsize MG5}}(C_1,C_2)$. 

The main $t\bar t$ production channel at LHC is given by gluon fusion and the associated Feynman diagrams are those  in Fig.~\ref{diagrams}. Other sub-leading channels are given by quark-antiquark annihilation, whose diagrams are depicted in Fig.~\ref{diagrams2}. MG5 computes the square of the amplitude for each single channel and then convolutes the result with the probability distribution functions (pdf) of the partons inside the proton in order to obtain the total $pp\to t\bar t$ production cross section. The default set of pdf used is {\sc CTEQ6L1}.

The partonic level result thus obtained can be compared with the partonic experimental cross section that is extracted by the experimental collaborations from the fully hadronized cross section---which is what is actually measured at the colliders.

We compute, using MG5,  $\sigma_{\mbox{\scriptsize MG5}}(C_1,C_2)$ for three different values of the center of mass energy ($7$, $8$ and $14$ TeV), varying the absolute values of both $C_1$ and $C_2$ in a range that goes from $0$ to $0.1$. These different values of $\sigma_{\mbox{\scriptsize MG5}}(C_1,C_2)$ will be used to obtain limits on the coefficients $C_1$ and $C_2$ by comparing the MG5 computation with the  measured cross section at the center-of-mass (CM) energy $\sqrt{s} = 7$ and $8$ TeV and the expected result at $14$ TeV, as discussed in the next section.

By proceeding in the same way, we have also computed the $t\bar t$ production cross section at the Tevatron and compared it with the measured cross section at the CM energy $\sqrt{s} = 1.98$ TeV. In this case, the main $t\bar t$ production channel is given by quark-antiquark annihilation and the associated Feynman diagrams are those depicted in Fig.~\ref{diagrams2}.
As we shall see, in this case we obtain a particular stringent  bound on $C_1$. 

\subsection{Statistical analysis}

The quantity used to obtain $95\%$ confidence level (CL) limits on the coefficients $C_1$ and $C_2$ is the  cross section $\Delta\sigma_{\mbox{\scriptsize exp}}$, which is defined to be the difference between the  central value of the measured cross section $\bar \sigma_{\mbox{\scriptsize exp}}$ and that of the  SM theoretical value $\bar \sigma_{\mbox{\scriptsize th}}$:
\be
\Delta \sigma_{\mbox{\scriptsize exp}} = \bar \sigma_{\mbox{\scriptsize exp}} - \bar \sigma_{\mbox{\scriptsize th}} \, .
\ee
 The uncertainty is given by summing in quadrature  the respective uncertainties:
\be
\sqrt{(\delta\sigma_{\mbox{\scriptsize exp}})^2 + (\delta\sigma_{\mbox{\scriptsize th}})^2}\,.
\ee

Using the cross sections $\sigma_{\mbox{\scriptsize MG5}}(C_1,C_2)$ calculated with MG5 we compute the value of the cross section coming from new physics $\Delta \sigma_{\mbox{\scriptsize MG5}}(C_1,C_2)$ as
\be
\Delta \sigma_{\mbox{\scriptsize MG5}}(C_1,C_2) =  \sigma_{\mbox{\scriptsize MG5}}(C_1,C_2) - \sigma_{\mbox{\scriptsize MG5}}(0,0)\,.
\ee

The quantity $\Delta \sigma_{\mbox{\scriptsize MG5}}(C_1,C_2)$ represents the contribution to the cross section coming from the  interference between SM leading order and new physics diagrams. Terms coming from the interference between new physics and higher order QCD diagrams are not included in this approximation.

Values of $C_1$ and $C_2$ for which
$\Delta \sigma_{\mbox{\scriptsize MG5}}(C_1,C_2)$ is more
than two standard deviations from $\Delta \sigma_{\mbox{\scriptsize exp}}$, namely
\be
\Delta \sigma_{\mbox{\scriptsize MG5}}(C_1,C_2) > \Delta \sigma_{\mbox{\scriptsize exp}} + 2 \, \sqrt{(\delta\sigma_{\mbox{\scriptsize exp}})^2 + (\delta\sigma_{\mbox{\scriptsize th}})^2} \, , \label{delta1}
\ee
or
\be
\Delta \sigma_{\mbox{\scriptsize MG5}}(C_1,C_2) < \Delta \sigma_{\mbox{\scriptsize exp}} - 2 \, \sqrt{(\delta\sigma_{\mbox{\scriptsize exp}})^2 + (\delta\sigma_{\mbox{\scriptsize th}})^2} \, , \label{delta2}
\ee
can be considered excluded at 95\% CL.

\section{Results}

\subsection{SM cross section}
\label{sec:xsecs}

A typical one-loop radiative correction to the vertices gives a contribution to the top quark radius  $O(\alpha_s/2 \pi m_t^2)$. Because the effect of compositeness is of the same order, the SM theoretical amplitude $\sigma_{\mbox{\scriptsize th}}$ used for obtaining the exclusion limits, as explained in the previous section, must contain at least NLO contributions which include these corrections as well as those coming from initial and final state radiation.

Production of top quark pairs at hadron colliders is a challenging computation that has been pursued for many years and is now available  at the next-to-next-to-leading order (NNLO)~\cite{Czakon:2013goa}. In addition, the soft gluon re-summation for the same process is known  at the next-to-next-to-leading logs (NNLL) order necessary for the matching~\cite{Cacciari:2011hy}. Following \cite{Czakon:2013goa} we take, for a top-quark mass of 172.5 GeV, the  following values for the cross section at the LHC for, respectively,  the CM energy $\sqrt{s} = 7$ and $8$:
\be
\sigma_{\mbox{\scriptsize th}} (pp\rightarrow t\bar t ) = 
\left\{  \begin{array}{cc}   
176.25 ^{+4.6}_{-5.9}\; 
\mbox{(scale)} \; ^{+4.8}_{-4.9} \; \mbox{(pdf)} \; \mbox{pb} & \mbox{(LHC@7)} \\ 
251.68 ^{+6.4}_{-8.6}\; 
\mbox{(scale)} \; ^{+6.3}_{-6.5} \; \mbox{(pdf)} \; \mbox{pb} & \mbox{(LHC@8)}                                       
\end{array}  \right. , \label{theory}
\ee
where the first uncertainty is due to the residual scale dependence and the second to the pdf of the partons which are taken from the {\sc MSTW200nnlo68cl} set~\cite{Martin:2009bu}. At the Tevatron, for  a CM energy $\sqrt{s}=1.98$ TeV, the same reference~\cite{Czakon:2013goa} gives
\be\label{theoryTeva}
\sigma_{\mbox{\scriptsize th}} (p\bar p\rightarrow t\bar t ) = 
 7.35 ^{+0.11}_{-0.21}\; 
\mbox{(scale)} \; ^{+0.17}_{-0.12} \; \mbox{(pdf)} \; \mbox{pb}  \mbox{ (Tevatron)}                                     \, .
\ee
The size of the overall uncertainty of these results---summing the  square of the two errors---is   a substantial improvement with respect to the NLO result.

\subsection{Current data and bounds: LHC and Tevatron}
\label{sec:limits} 

The cross section for the production of top quark pairs has been measured at LHC and Tevatron for their respective energy range.

%%%%%%%%%%%%%%%%%%%
\begin{figure}[ht!]
\begin{center}
\includegraphics[width=3.0in]{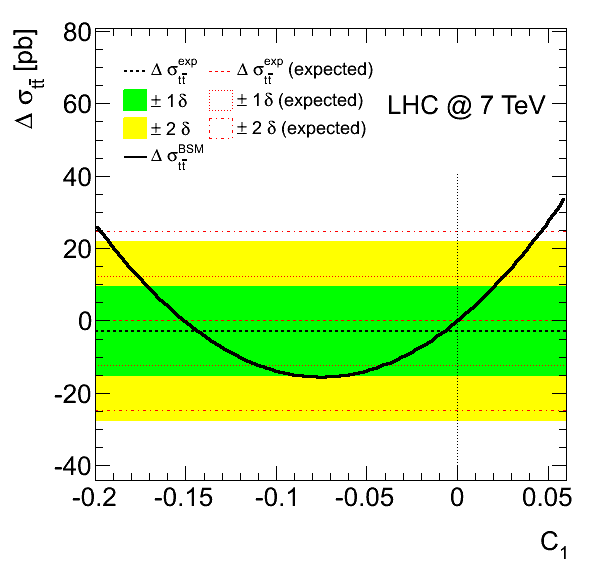}
\includegraphics[width=3.0in]{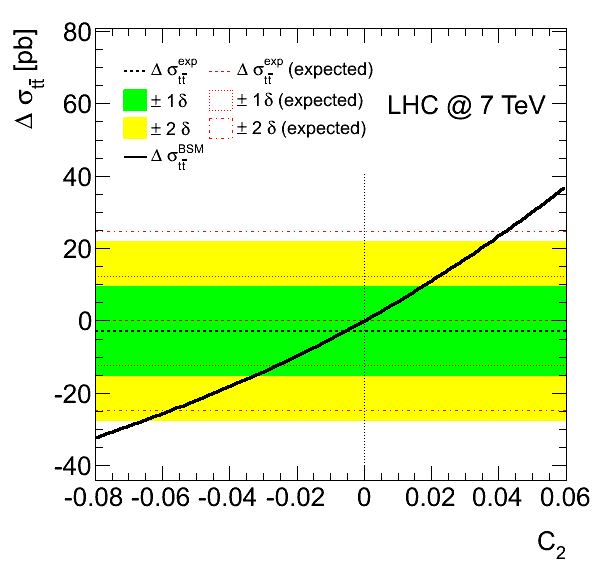}
\caption{\small 
\label{LHC7}Constraints on the coefficients $C_1$ and $C_2$ from data at the LHC at 
 $\sqrt{s}=7$ TeV. On the left: the limit on $C_1$ with $C_2=0$. On the right, the limit on $C_2$ with $C_1=0$. The horizontal dashed black  line represents the experimental central value. The  yellow (green) band represents 2(1)$\sigma$ uncertainties. The red lines are the expected limits at 1 and 2$\sigma$ level.  The thick black line is the cross section in the presence of the new operators at sampled values of the coefficients $C_i$.}
\end{center}
\end{figure}
%%%%%%%%%%%%%%%%%%%%%%

The best current measurements at the LHC of the cross section $\sigma_{\mbox{\scriptsize exp}} (pp\rightarrow \bar t t)$ combining the various channels at the CM energy of  $\sqrt{s}=7$ TeV, for a top-quark mass of 172.5 GeV, is
\be
\sigma_{\mbox{\scriptsize exp}} (pp\rightarrow t\bar t ) = 
\left\{  \begin{array}{ll} 
177.0 \pm 3 \, \mbox{(stat)} ^{+8}_{-7} \, \mbox{(syst)} \pm 7 \, \mbox{(lumi)}  \, \mbox{pb}   &\mbox{(ATLAS)}  \nn \\
  165.8 \pm 2.2\, \mbox{(stat)}  \pm 10.6\, \mbox{(syst)} \pm 7.8 \, \mbox{(lumi)}  \: \mbox{pb}  & \mbox{(CMS)} 
  \end{array} \right.
\ee
for, respectively ATLAS~\cite{ATLAS} and CMS~\cite{CMS}. 

A combination of ATLAS and CMS results is  available~\cite{ATLAS:2012dpa} for an integrated luminosity of up to 1.1 fb$^{-1}$:
\be
\sigma_{\mbox{\scriptsize exp}} (pp\rightarrow \bar t t) = 173.3 \pm 10.1 \: \mbox{pb}\qquad\,\mbox{(LHC@7)}\,,
\ee
with an overall uncertainty of 5.8\% which we will use in our analysis to set the limits.

Fig.~\ref{LHC7} show the limits coming from LHC@7 on the coefficients $C_1$ and $C_2$ obtained by means of the above experimental result and the theoretical computation in \eq{theory}. 

In Fig.~\ref{LHC7}, as well as in the following figures, the black line with dots represents the cross section $\Delta \sigma_{\rm MG5}$ which includes the contributions of the operators $\hat O_1$ and $\hat O_2$. The yellow (green) bands represents the cross section $\Delta \sigma_{\mbox{\scriptsize exp}}$ with its error at $2(1) \sigma$ level. Finally, the horizontal red lines represent the expected limits (at the 1 and 2$\sigma$ level)  which are obtained by  identifying the central value of the experimental data $\bar \sigma_{\mbox{\scriptsize exp}}$ with the central value of the theoretical prediction $\bar \sigma_{\mbox{\scriptsize th}}$.

%%%%%%%%%%%%%%%%%%%
\begin{figure}[ht!]
\begin{center}
\includegraphics[width=3.0in]{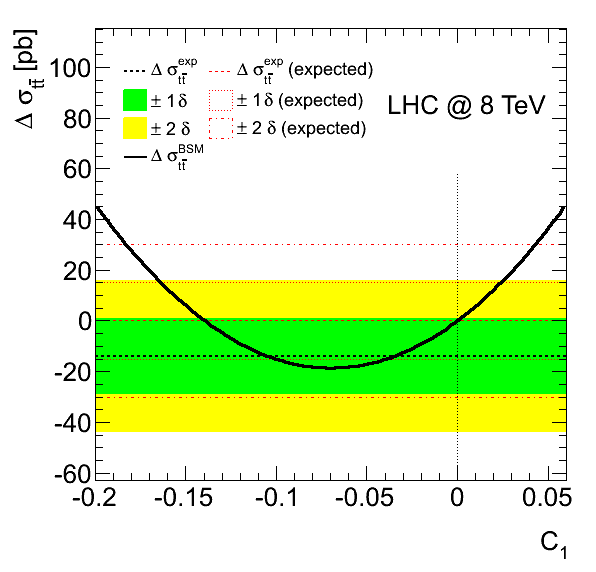}
\includegraphics[width=3.0in]{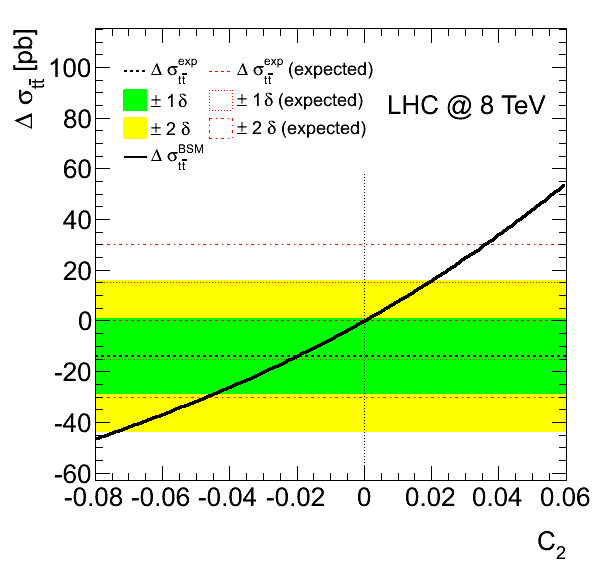}
\caption{\small Constraints on the coefficients $C_1$ and $C_2$ from data at the LHC at 
 $\sqrt{s}=8$ TeV. On the left, the limit on $C_1$ with $C_2=0$. On the right, the limit on $C_2$ with $C_1=0$. The horizontal dashed black line represents the experimental central value. The  yellow (green) band represents 2(1)$\sigma$ uncertainties. The red lines are the expected limits at 1 and 2$\sigma$ level.  The black line is the cross section in the presence of the new operators at sampled values of the coefficients $C_i$.
 \label{LHC8}}
\end{center}
\end{figure}
%%%%%%%%%%%%%%%%%%%%%%

The best current measurements of the cross section $\sigma_{\mbox{\scriptsize exp}} (pp\rightarrow \bar t t)$ at the CM energy of  $\sqrt{s}=8$ TeV for a top-quark mass of 172.5 GeV is
\be
% NEW, fixed Eq.18:
\sigma_{\mbox{\scriptsize exp}} (pp\rightarrow t\bar t ) =
\left\{  \begin{array}{ll} 237 \pm 1.7 \; \mbox{(stat)} \pm 7.4 \;
\mbox{(syst)} \pm 7.4\; \mbox{(lumi)} \pm 4.0\; \mbox{(beam energy)}
\: \mbox{pb} & \mbox{(ATLAS \cite{ATLAS8})} \\
 227 \pm 3\; \mbox{(stat)} \pm 11\; \mbox{(syst)} \pm 10 \;
\mbox{(lumi)}  \: \mbox{pb}&    \mbox{(CMS \cite{CMS8})}
 \end{array} \right.
\ee
for, respectively, an integrated luminosity of 5.8 fb$^{-1}$ at ATLAS
and 2.4 fb$^{-1}$ at CMS,
in both cases considering events with dilepton final states.

Lacking a combined value, we consider the experimental value of ATLAS~\cite{ATLAS8}, which has smaller uncertainties, to set the limits; the theoretical value is taken from \eq{theory}. Fig.~\ref{LHC8}  shows the result in this case. Notice that  improved limits with respect to the previous ones at $\sqrt{s}=7$ TeV are mainly due to the lower  central value of the experimental data. This is made clear by  the comparison with the expected CL (the red horizontal lines in   Fig.~\ref{LHC8} ). 

%%%%%%%%%%%%%%%%%%%
\begin{figure}[ht!]
\begin{center}
\includegraphics[width=3.0in]{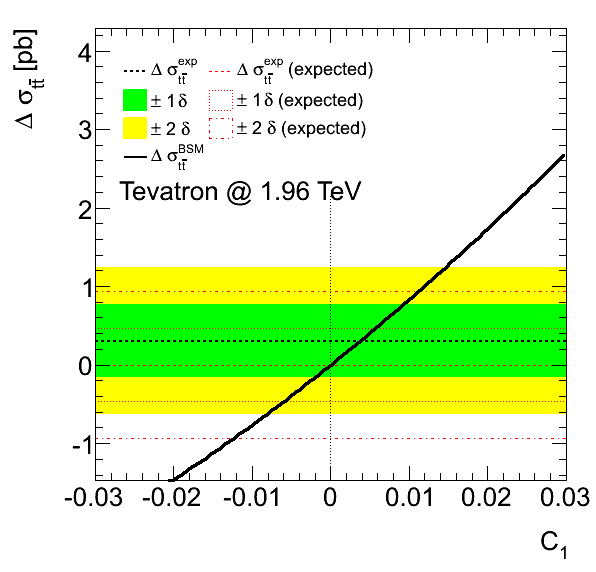}
\includegraphics[width=3.0in]{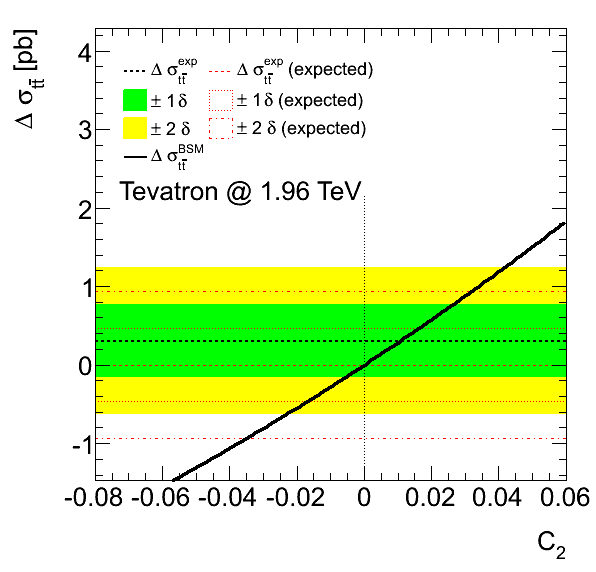}
\caption{\small 
\label{Teva} Constraints on the coefficients $C_1$ and $C_2$ from data at the Tevatron at $\sqrt{s}=1.96$ TeV. On the left, limit on $C_1$ with $C_2=0$. On the right, limit on $C_2$ with $C_1=0$. The horizontal dashed black line represents the experimental central value. The  yellow (green) band represents 2(1)$\sigma$ uncertainties. The red lines are the expected limits at 1 and 2$\sigma$ level.  The black line is the cross section in the presence of the new operators  at sampled values of the coefficients $C_i$.}
\end{center}
\end{figure}
%%%%%%%%%%%%%%%%%%%%%%

Combined data from CDF and D0 at  the Tevatron~\cite{Teva} give the following cross section at the CM energy of $\sqrt{s}=1.96$ TeV up to an integrated luminosity of 8.8  fb$^{-1}$:
\be
\sigma_{\mbox{\scriptsize exp}} (p\bar p\rightarrow \bar t t) = 7.65 \pm 0.42 \: \mbox{pb}
\qquad\,\mbox{(Tevatron)}\, .
\ee
Fig.~\ref{Teva} show the limits coming from Tevatron on the coefficients $C_1$ and $C_2$ we obtain by means of the above experimental result and the theoretical computation in \eq{theoryTeva}. The data from the Tevatron are particularly stringent in the case of the operator $\hat O_1$ because of the kinematical configuration that prefers the $qq\rightarrow t\bar t$ channel which is, in turn, most sensitive to that operator.

\subsection{Future bounds: LHC at $\sqrt{s}=14$ TeV}

%%%%%%%%%%%%%%%%%%%
\begin{figure}[ht!]
\begin{center}
\includegraphics[width=3.0in]{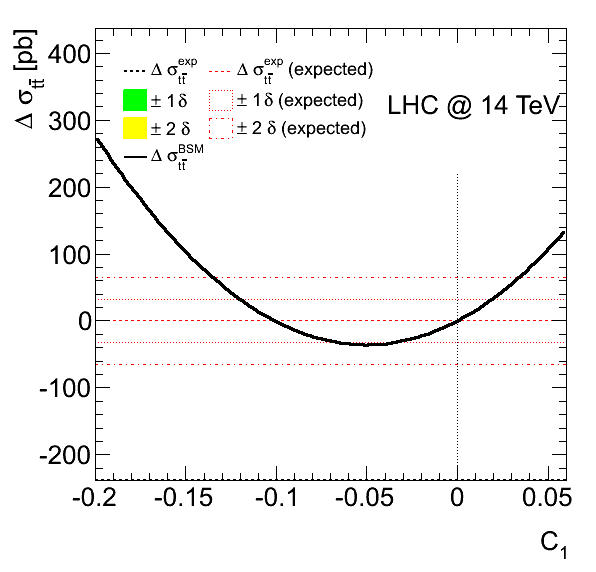}
\includegraphics[width=3.0in]{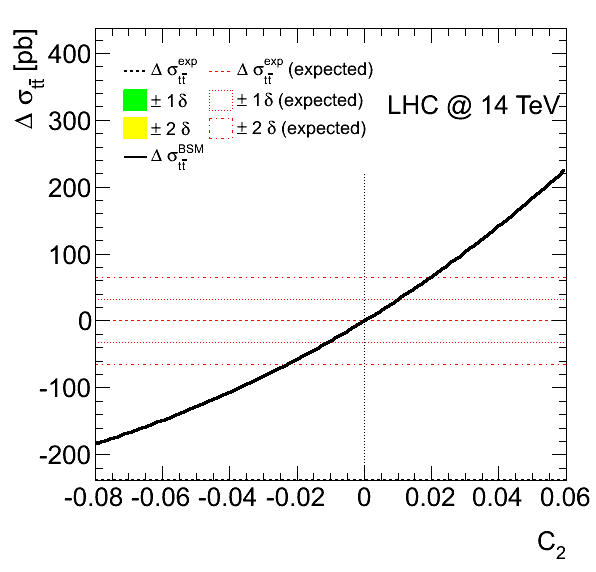}
\caption{\small Possible constraints on the coefficients $C_1$ and $C_2$ from data at the LHC at 
 $\sqrt{s}=14$ TeV. On the left, limit on $C_1$ with $C_2=0$. On the right, limit on $C_2$ with $C_1=0$. The horizontal dashed red line represents the experimental central value. The other red lines are the expected limits at  1 and 2$\sigma$ level. The black line is the cross section in the presence of the new operators  at sampled values of the coefficients $C_i$.
\label{LHC14}}
\end{center}
\end{figure}
%%%%%%%%%%%%%%%%%%%%%%

If we assume that the experimental uncertainty will remain around 5\%---it is difficult to imagine doing better than this---we can plot the expected limits at the LHC when the CM energy  will be $\sqrt{s}=14$ TeV by fixing the  experimental central value to coincide with the theoretical cross section:
\be
\sigma_{\mbox{\scriptsize th}} (p\bar p\rightarrow t\bar t ) = 
953.6  ^{+22.7}_{-33.9} \mbox{(scale)}  ^{+16.2}_{ -17.8} \mbox{(pdf)} \; \mbox{pb} \;\mbox{(LHC@14)~\cite{Czakon:2013goa}} \, .
\ee

  As depicted in Fig.~\ref{LHC14}, the increase in energy  improves the limits with respect to what was to be expected at the LHC at lower CM energies. However,   the study at 14 TeV does not  modify in a significative manner the overall limits  because the actual experimental value at 8 TeV turned out lower than the theoretical value and therefore yielded a better than expected limit.

%%%%%%%%%%%%%%%%%%%%%%%%%%%%%%%%%%%%%%%%
\subsection{Differential cross section}

%%%%%%%%%%%%%%%%%%%
\begin{figure}[ht!]
\begin{center}
\includegraphics[width=3.5in]{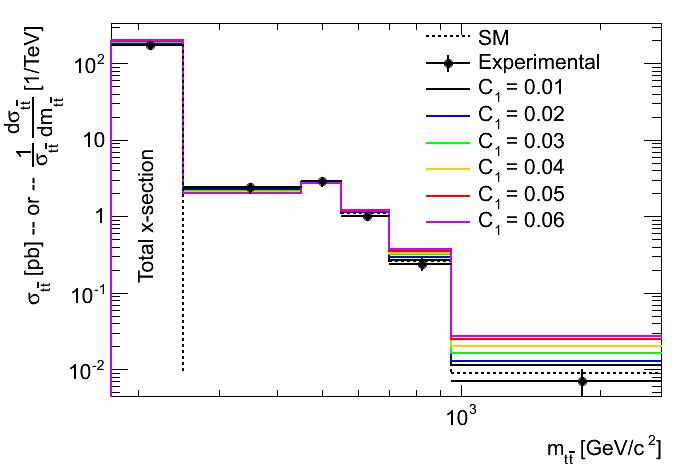}
\includegraphics[width=3.5in]{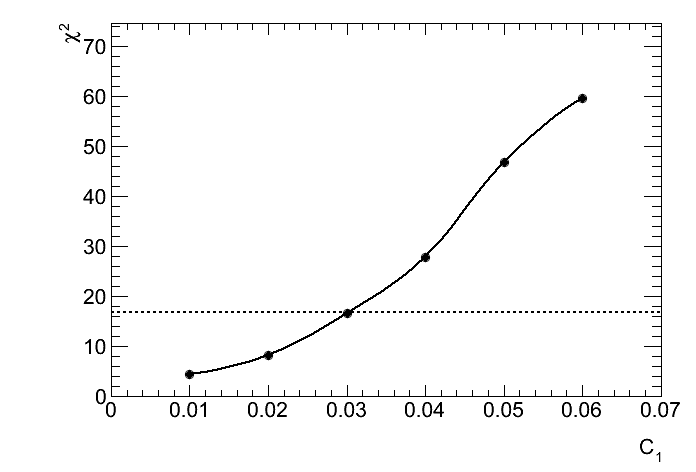}
\caption{\small Left side: Differential cross section in bins of $m_{t\bar t}$ for LHC at 7 TeV. In the first bin, the total cross section. In color, the changes bin by bin for different values of the coefficient $C_1$. Right side: Limit on the coefficient $C_1$ (see the text).
\label{diff}}
\end{center}
\end{figure}
%%%%%%%%%%%%%%%%%%%%%%

Even though the differential cross section for $t\bar t$ production contains, in principle, extra  information that can be used to set limits on the compositeness of the top quark,  we find that the current experimental uncertainties are too large to  significantly improve the best limits we found by considering the total cross section.

The best case occurs for the cross section as a function of the invariant mass $m_{t\bar t}$ for the LHC with data at $\sqrt{s} = 7$ TeV~\cite{diff_exp} and for the coefficient $C_1$. Fig.~\ref{diff} plots this cross section in bins and shows the experimental value with its error and the variation for different values of the coefficient $C_1$. 

To set limits combining the total cross section $\sigma_{tot} = \sigma_{t\bar{t}}$ 
and the relative differential cross section 
$\sigma_{dif,i} = (1/\sigma_{t\bar{t}})\, d\sigma_{t\bar{t}}/dm_{t\bar{t}} $ in each bin $i$, 
we evaluated a $\chi^2$ function as:
\begin{equation}
\chi^2 =           \left[ \frac{\sigma_{tot }^{exp} -  \sigma_{tot }^{th}}{\delta(\sigma_{tot })} \right]^2
 + \sum_{i=0}^{N_{bins}}  \left[  \frac{\sigma_{dif,i}^{exp} - \sigma_{dif,i}^{th}}{\delta(\sigma_{dif,i})} \right]^2 \, ,
\end{equation}
where $\delta(\sigma)$  refers to the squared sum of the experimental 
and the theoretical uncertainties on the respective $\sigma_{tot}$  and $\sigma_{dif,i}$, 
and the suffix $th$ refer to the theoretical prediction.
This theoretical prediction is given by the NLO+NLL calculation~\cite{diff_th} plus the 
contribution from new physics depending on the value of $C_1$, 
evaluated generating events with {\sc Madgraph} and then {\sc Pythia}~\cite{pythia}. 

This quantity $\chi^2$ is evaluated for each of the considered values of $C_1$ 
and compared with a distribution of $\chi^2$ values obtained generating $10^3$ pseudo-experiments 
allowing the measured values to fluctuate following a gaussian distribution $G(\sigma,\delta(\sigma))$. 
A value of $C_1$ was considered excluded at 95\% C.L. if less than 5\% of the pseudo-experiments 
resulted in a $\chi^2$ value larger than the one for the given value of $C_1$. 

To get an approximate 95\% C.L. exclusion limit on $C_1$, 
the obtained values of $\chi^2$ as a function of $C_1$ were interpolated as shown on the right side of Fig.~\ref{diff}.
The dashed horizontal line represents the value $\hat{\chi}^2$ for which $\chi^2<\hat{\chi}^2$ 
in 95\% of the pseudo-experiments.
The limit we find is $C_1 < 0.03$, which is an improvement with respect to what found from data on the LHC total cross section at the same CM energy, but still less stringent than that found from the Tevatron data for the total cross section. 

Direct analysis of the Tevatron data~\cite{diff_teva} and the LHC at $\sqrt{s}=8$ TeV~\cite{diff_8} yield weaker limits. The effect of the coefficient $C_2$ on the cross section distributions is negligible because of the experimental uncertainties in the different intervals of $m_{t\bar{t}}$.

%%%%%%%%%%%%%%%%%%%%%%%%%%%%%%%%%%%%%%%%
\subsection{Spin correlations: LHC at $\sqrt{s}=7$ TeV}

Independent observables useful in setting further limits on the top quark structure involve spin correlation in $t\bar{t}$ events~\cite{Bernreuther:2013aga}. 
Spin correlations of pair-produced top quarks can be extracted by analysing the angular distributions
of the top-quark decay products in $t \to W b$ followed by the leptonic decay of the $W$ boson $W\to l \nu$.
If no acceptance cuts are applied and the spin-analysing power is effectively $100\%$, we have the following  form of the double angular distribution~\cite{Bernreuther:2013aga}
\be \label{cosdistr}
\frac{1}{\sigma}\frac{d^2\sigma}{d\cos\theta_1d\cos\theta_2}=\frac{1}{4}(1+B_1\cos\theta_1+B_2\cos\theta_2-C_h\cos\theta_1\cos\theta_2)\,,
\ee
where $B_1$, $B_2$ and $C_h$ are coefficients.

%%%%%%%%%%%%%%%%%%%
\begin{figure}[h!]
\begin{center}
\includegraphics[width=3.0in]{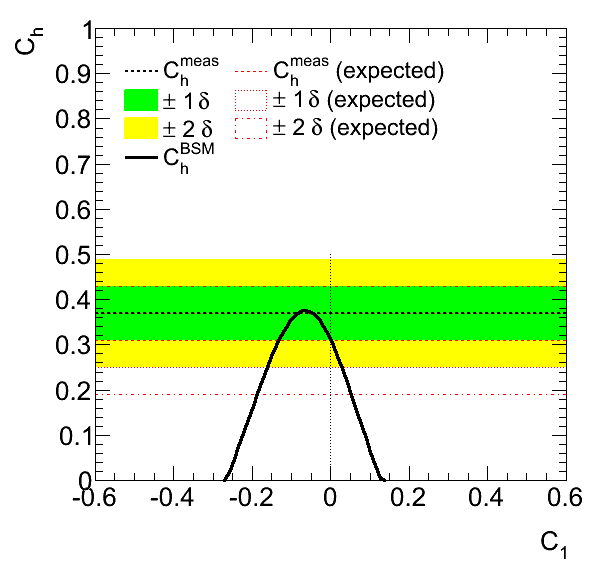}
\includegraphics[width=3.0in]{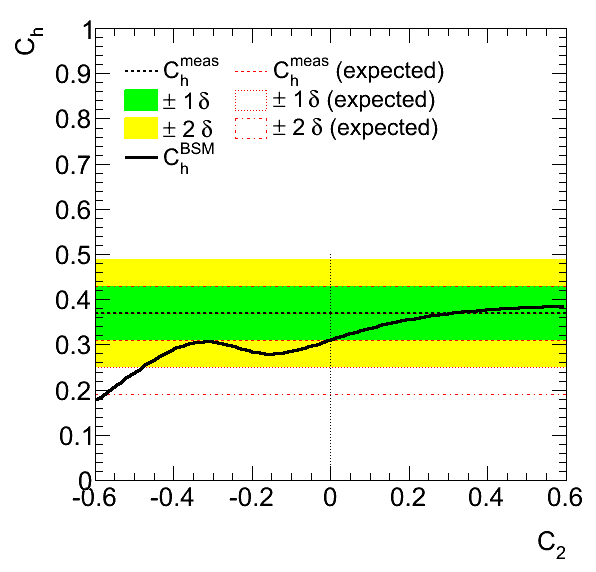}
\caption{\small 
\label{spin} Constraints on the coefficients $C_1$ and $C_2$ from data on  spin correlations at the LHC at $\sqrt{s}=7$ TeV. On the left, limit on $C_1$ with $C_2=0$. On the right, limit on $C_2$ with $C_1=0$. The horizontal dashed black line represents the experimental central value. The  yellow (green) band represents 2(1)$\sigma$ uncertainties. The red lines are the expected limits at 1 and 2$\sigma$ level.  The black line is the asymmetry $C_h$ computed in the presence of the new operators as a function of the coefficients $C_i$.}
\end{center}
\end{figure}
%%%%%%%%%%%%%%%%%%%%%%

Here we consider the angular distribution in the helicity basis, in which the quantization axes are taken to be the $t$ and $\bar t$ directions of flight in the $t\bar t$ zero-momentum frame. Therefore $\theta_1(\theta_2)$ represents the angle between the direction of flight of $l^+(l^-)$ in the $t(\bar t)$ rest frame and the $t(\bar t)$ direction of flight in the $t\bar t$  zero-momentum frame. In this case the spin correlation coefficient $C_h$ is given by
\be \label{chel}
C_h = \frac{N(\uparrow\uparrow) + N(\downarrow\downarrow)- N(\downarrow\uparrow) -
N(\uparrow\downarrow)}{N(\uparrow\uparrow) + N(\downarrow\downarrow)+N(\downarrow\uparrow) +
N(\uparrow\downarrow)} \, 
\ee 
where $N(\uparrow\uparrow)+ N(\downarrow\downarrow)$ represents the number of events where the top and antitop quark spin projections are
parallel, and $N(\downarrow\uparrow) +N(\uparrow\downarrow)$ is the number of events where they are anti-parallel with respect to the chosen quantization axes.

For the experimental analysis it is more convenient to use the one-dimensional distributions of the product of the cosines 
$O_h\equiv \cos\theta_1\cdot \cos\theta_2$ and define an asymmetry $A_h$ which, in the absence of acceptance cuts, is determined by~\cite{Bernreuther:2013aga}
\be
\label{eq:Ah}
A_h=  \frac{N(O_h>0) - N(O_h<0)}{N(O_h>0) + N(O_h<0)}=-\frac{C_h}{4}\,.
\ee

The most precise experimental measurement of $C_h$ at LHC is obtained by ATLAS studying at $\sqrt{s} = 7$ TeV the angular separation $\Delta\phi$ between the charged leptons in dileptonic $t\bar{t}$ events ~\cite{TheATLAScollaboration:2013gja}
\be
C_h^{meas} = 0.37 \pm 0.06 (stat+syst)\, .
\ee
This value has to be compared with the next-to-leading order SM prediction
\be\label{chsm}
C_h^{NLO} = 0.31,
\ee
calculated in~\cite{Bernreuther:2013aga} including QCD
and electroweak corrections to $t\bar{t}$ production and decay.

The spin correlation coefficient $C_h$ as a function of the effective operators coefficients is obtained by means of MG5, generating  $10^5$ events for the process $pp \rightarrow t\bar{t} \rightarrow b \ell^+ \nu \bar{b} \ell^- \bar{\nu}$, for different values of $C_1$ and $C_2$, and computing the asymmetry $A_h$ defined in \eq{eq:Ah}.

MG5 gives a tree level prediction for the total cross section $\sigma(pp \rightarrow t\bar{t} \rightarrow b \ell^+ \nu \bar{b} \ell^- \bar{\nu})$ that we denote by $\sigma^{MG5}(C_1,C_2)$ and therefore also the derived asymmetry is a LO result that we denote it by $A_h^{MG5}(C_1,C_2)$.
The tree level prediction for the asymmetry is then corrected in order to take into account the SM NLO effects by computing
\be
A_h(C_1,C_2) = 
  \frac{A_h^{NLO}\cdot \sigma^{NLO}   +  A_h^{MG5}(C_1,C_2)\cdot\sigma^{MG5}(C_1,C_2)  -  A_h^{MG5}(0,0)\cdot \sigma^{MG5}(0,0) }
       {\sigma^{NLO}  +  \sigma^{MG5}(C_1,C_2)  -  \sigma^{MG5}(0,0) }\,,
\ee
where $\sigma^{NLO}$ is obtained by multiplying the NNLO theoretical cross section for $t\bar t$ production in \cite{Czakon:2013goa} by the leptonic decay branching ratio and $A_h^{NLO}$ is computed using \eq{eq:Ah} from \eq{chsm} .

Fig.~\ref{spin} show the limits on the coefficients $C_1$ and $C_2$ coming from the spin correlations at the LHC at $\sqrt{s}=7$ TeV. The black line represents the correlation coefficient  computed including the contributions of the operators $\hat O_1$ and $\hat O_2$ as described above. The yellow (green) bands represents the measured $C_h$ with its error at 2(1)$\sigma$ level. We can see from Fig.~\ref{spin} that differently to $C_1$, the coefficient $C_2$ is unbounded for positive values.

%%%%%%%%%%%%%%%%%%%%%%%%%%%%%%%%%%%%%%%%
\subsection{Combined limits}

\begin{figure}[ht!]
\begin{center}
\includegraphics[width=3.5in]{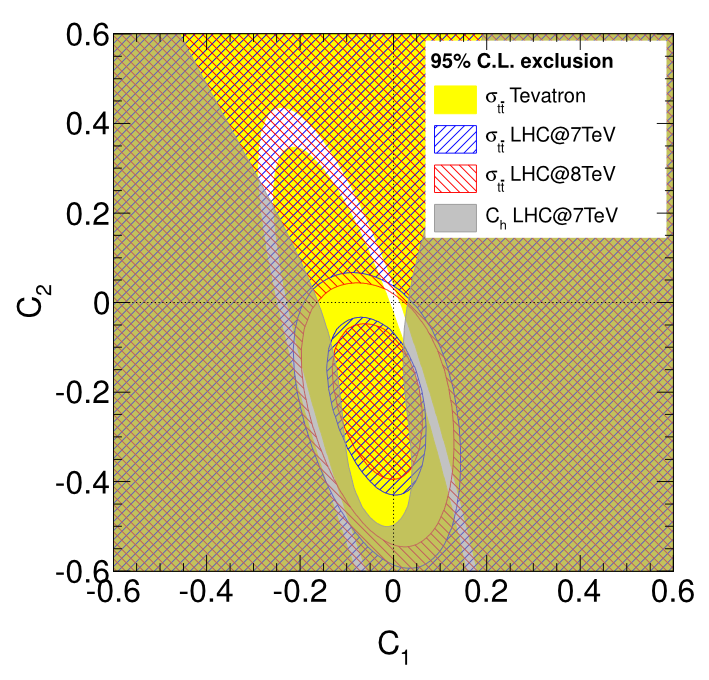}
\includegraphics[width=3.5in]{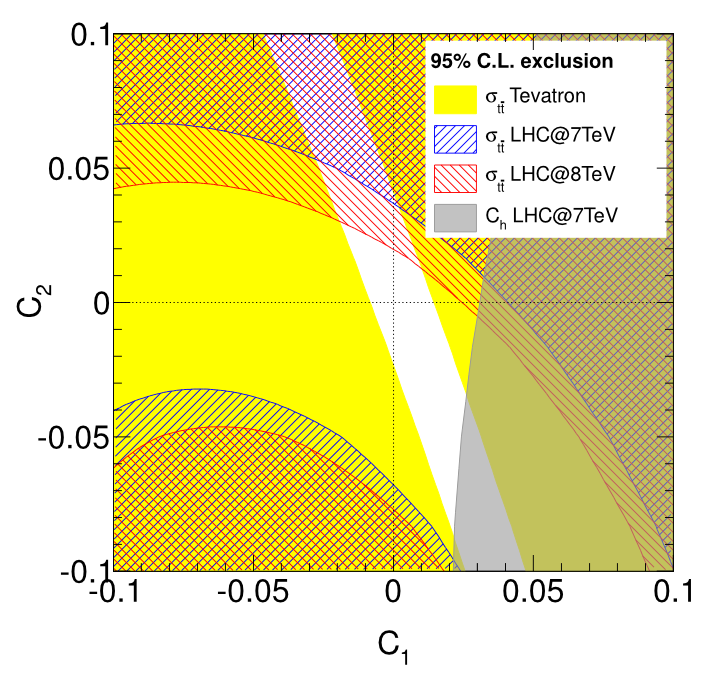}
\caption{\small Combined limits on the coefficients $C_1$ and $C_2$ from data at the LHC ($\sqrt{s}=8$ and 7 TeV) and Tevatron ($\sqrt{s}=1.96$ TeV). Values in the regions, respectively, in hatched blue and hatched red from LHC, yellow from Tevatron and gray from spin correlations at the LHC   are excluded at 95\% CL. The allowed region corresponds to the white area. On the right a zoomed plot around the allowed values.
 \label{combo}}
\end{center}
\end{figure}

The bounds on the individual coefficients $C_1$ and $C_2$ can also be computed  when both operators are simultaneously present. In this case, there is   an important modulation in the effect of the new operators. The combined limits coming from the data at the LHC at the two CM energies considered and those at the Tevatron are shown in Fig.~\ref{combo}. In the same plot, also the limits coming from the spin correlations are shown.

We can see that without the data on the  spin correlations the allowed region  is rather large. Other analysis~\cite{Lillie:2007hd,kamenik,hioki,Englert:2012by,Degrande:2010kt} found a smaller region because they either  kept  only the leading order contribution of the operators  or did not combine the limits on $C_2$ with those on $C_1$. When the full contribution of these operators is included for both of them, we need the limits from data on the spin correlations in order to exclude the larger values and obtain relevant limits.

%%%%%%%%%%%%%%%%%%%
%%%%%%%%%%%%%%%%%%%%%%

%\clearpage
\section{Discussion}
\label{sec:discussion} 
%%%%%%%%%%%%%%%%%%%%%%%
\begin{table}[ht!]
\begin{center}
\caption{Limits on the coefficients $C_1$ and $C_2$ when they are varied independently of each other and simultaneously (last column). The starred numbers for LHC@14 are the expected limits.}
\label{limits}
\vspace{0.2cm}
%\begin{ruledtabular}
\begin{tabular}{|c|c|c|c|c|}
\quad  Tevatron \quad &\quad  LHC@7 \quad  & \quad LHC@8 \quad & \quad LHC@14 \quad & combined\cr
\hline
 $- 0.008 \leq C_1 \leq $ 0.015 & $-0.193 \leq C_1 \leq $ 0.042 & $ -0.165 \leq C_1 \leq $ 0.025 &  $-0.135^\star \leq C_1 \leq $ 0.034$^\star$ &  $-0.019 \leq C_1 \leq 0.026$ \cr
\hline
$-0.023 \leq  C_2  \leq   0.042$ & $-0.066 \leq  C_2  \leq  0.038$ & $-0.074 \leq  C_2  \leq  0.020$ & $-0.023^\star \leq  C_2  \leq  0.020^\star$ & $-0.087 \leq  C_2  \leq  0.031$ \cr
%\hline
\end{tabular}
%\end{ruledtabular}
\end{center}
\end{table}
%%%%%%%%%%%%%%%%%%%%%%%%%

We have collected in Table~\ref{limits} the limits for the two coefficients $C_1$ and $C_2$ when they are varied simultaneously and independently of each other for each of the data sets. Pending results from the LHC at $\sqrt{s} = 14$ TeV, the Tevatron gives the most stringent bound on the operator $\hat O_1$. The LHC at $\sqrt{s} = 8$ TeV and the Tevatron give the best bounds on the operator $\hat O_2$. In combining the limits, all data sets are important and those from spin correlations are essential in removing  large values for the coefficients $C_1$ and $C_2$ which  cannot be ruled out by the total cross section data because of cancellations between the two contributions.

\subsection{The size of the top quark}

Deviations from the point-like behavior of a particle are usually expressed in terms of charge radius and anomalous magnetic moment. These quantities have been defined in \eq{R} and \eq{MU}.  In the case of the top quark the dominant probe is charged under the $SU(3)$ group of strong interactions. Since there is no definite boundary, the size is usually discussed in terms of root mean squared (RMS) radius.

We should first identify the size that comes from radiative corrections. This part is a departure from the point-like behavior due to the cloud of virtual states surrounding any particle in quantum field theory. For the top quark, the leading radiative contribution to the squared mean radius can be computed in perturbation theory  to one-loop order in QCD and it is in general of order $(\alpha_s/2\pi) 1/m_t^2$. Accordingly, this contribution to the the RMS radius  is  about $10^{-5}$ fm. EW interactions give an additional (smaller) contribution  which can be neglected.

 We can obtain an estimate of the fraction of the radius and moment of the composite top quark which is not part of the radiative corrections in terms of the effective operator coefficients. Using the formulas in \eq{f12} to compute the form factors of \eq{gem}, we have that, using \eq{rm},
\be
\langle \vec r^2 \rangle =  \frac{6}{4} \frac{4 m_t\, C_1 + 2\, \upsilon\, C_2 }{m_t^3}  \quad \mbox{and} \quad \mu  =  1 +2\, \frac{\upsilon}{m_t} C_2 \, ,
\ee
with $2 \upsilon C_2/m_t$ equal to the anomalous component of the magnetic moment.
The best constrains on the coeficients $C_1$ and $C_2$ when taken independently
\be
-0.008 \leq C_1\leq 0.015 \qquad \mbox{and} \qquad   -0.023 \leq C_2 \leq 0.020
\ee
 correspond to a RMS radius
\be
\sqrt{\langle \vec r^2 \rangle} <  4.6 \times 10^{-4} \; \mbox{fm} \quad \mbox{(95\% CL)} \, , \label{radius}
\ee 
 and an anomalous magnetic moment 
\be
-0.046 < \mu -1 < 0.040 \quad \mbox{(95\% CL)} \, . \label{mu}
\ee
 A similar bound on the anomalous magnetic moment of the top quark has been reported in \cite{kamenik}. The bounds found in  \cite{Englert:2012by}  are weaker. At the time of writing this paper,  an apparently stronger bound was found in \cite{hioki} which however agrees with our limit if taken at the  68\% CL.

These limits become  weaker if the constraints are taken simultaneously for the two operators:
\be
\sqrt{\langle \vec r^2 \rangle} <  7.4 \times 10^{-4} \; \mbox{fm} \quad \mbox{and} \quad
-0.17 < \mu -1 < 0.062 \quad \mbox{(95\% CL)}
\ee

To put the above limits in prospective, let us consider the proton and the electron as the best known examples of, respectively, a composite  and (presumably) point-like particle.

The radiative part of the proton RMS radius   is about $10^{-2}$ fm. Its actual size,  as measured in electron-proton elastic scattering experiments, is much larger:   it is about 0.9 fm~\cite{Sick:2003gm}. The factor of 100 in the ratio of these two numbers is not far from what we have found in the case of the top quark. 

The electron, which is believed to be a point-like particle, has a limit on its RMS which does not originate in  radiative corrections that is not far from that in \eq{radius} for the top quark: it is  $10^{-5}$ fm~\cite{Bourilkov:2001pe}.  In other words,   the top quark does not seem to deviate from a point-like object down to a scale close to that of the electron. Notice that the bound in the electron case comes from an analysis of $e^+e^- \rightarrow e^+e^-$ while in our case that for the top quark comes from $\bar q q \rightarrow \bar t t$ rather than the more direct $\bar t t  \rightarrow \bar t t$ which would be probed by the process with four top-quarks in the final states. The limit on the part of the electron anomalous magnetic moment not accounted for by radiative corrections is very strong---because of its interaction with a classical magnetic field---and equal to $10^{-12}$ \cite{Gabrielse:2006gg}.

\subsection{Compositeness and physics beyond the SM}

The  most direct way to associate some compositeness scales $\Lambda_1$ and $\Lambda_2$ to the effect of the operators in \eq{O} is through the  identification 
\be
\frac{1}{\Lambda_1^2} = \frac{g_s |C_1|}{m_t^2} \quad \mbox{and} \quad
\frac{1}{\Lambda_2^2} =\frac{g_s  |C_2|}{2 m_t^2}\,. \label{l2}
\ee
The identification of $\Lambda_2$ in \eq{l2} is based on the full EW symmetry group for which the operator $\hat O_2$ must be considered of dimension 6.
In this way, it is simple  to translate the bounds on $C_1$ and $C_2$, obtained in the previous section, into limits on these two scales of compositeness. We have that
\be
\Lambda_{1} > 
\left\{  \begin{array}{cc} 1.3\; \mbox{TeV} & \mbox{(Tevatron)} \\ 
                                      0.4\; \mbox{TeV} & \mbox{(LHC@7)} \\ 
                                      0.4\; \mbox{TeV} & \mbox{(LHC@8)}
\end{array}  \right.
\quad \quad
\Lambda_{2} > 
\left\{  \begin{array}{cc} 1.1\; \mbox{TeV} & \mbox{(Tevatron)} \\ 
                                      0.9\; \mbox{TeV} & \mbox{(LHC@7)} \\ 
                                      0.8\; \mbox{TeV} & \mbox{(LHC@8)}
\end{array}  \right. 
\quad \mbox{(95\% CL)} \,,
\ee

These results can be compared with bounds coming from EW precision measurements~\cite{Ciuchini:2013pca}:  the scale of the four-fermion operators (for light quarks) is bounded to values  higher, depending on the sign and the procedure, than 6.6 or 9.5 TeV.  More generally, by using various other operators an overall  bound of $\Lambda > 17$ TeV is found.

While the above identification of the compositeness scale is straightforward---it plays the role of expansion parameter for the operators---its link to specific models is more indirect. 

In the following sections we discuss how to translate  bounds on $C_1$ and $C_2$ into limits on the parameters of two models of physics beyond the SM. In doing so, we must bear in mind that often numerical values, when used within a given model, are more orders of magnitude than precise numbers because both naive power counting and the QCD analogy may not be correct in a generic strongly interacting theory. 

\subsubsection{Contact interactions}

Contact interactions are usually introduced to parametrize generic models of compositeness~\cite{Eichten}. These have been traditionally described by the interaction
\be
\frac{2 \pi }{\Lambda^2_{\rm CI}} \;\bar \psi_{L}\gamma^\mu\psi_{L}\bar \psi_{L}\gamma_\mu\psi_{L}\, , \label{fermi}
\ee
where $\psi = (u\,\,d)^T$ and $\Lambda_{\rm CI}$ is identified as the contact interaction scale. The factor of $2\pi$ is conventional and suggested by a strongly interacting sector, the coupling of which is assumed to be $g^2 \simeq 2 \pi$. Under such an assumption, the effective four-fermion operator can be thought as generated by the exchange of a heavy resonance, which couples to the fermion with strength $g$, and that has been integrated out. The effect of four-fermion operators with light quarks in $\sigma(pp\rightarrow jj+X)$ was first discussed in \cite{Eichten} and more recently in~\cite{Bazzocchi:2011in,Domenech:2012ai}. 

Performing a field redefinition, using the equations of motion for the gluon field, it is possible to rewrite the operator $\hat O_1$ as a combination of four-fermion interaction terms:
\be\label{ff}
\hat O_1=\frac{g_s^2 C_1}{m_t^2} \;  \bar t \gamma_\mu T_A t\,\sum_q\bar q \gamma^\mu T^A q \,,
\ee
where the summation runs over all quark species. Assuming   contact interactions which are flavor universal, we can directly relate the contribution of the operator $\hat O_1$ to that in \eq{fermi}. By taking into account color factors, flavor and chirality multiplicity, we have the following identification:
\be
\Lambda_{\rm CI} = \sqrt{\frac{6 \pi}{g_s^2 C_1}} m_t\,.
\ee
The constraints on the coefficients $C_1$ therefore  applies to this scale and  gives
\be
\Lambda_{\rm  CI} > 
\left\{  \begin{array}{cc} 5.0\; \mbox{TeV} & \mbox{(Tevatron)} \\ 
                                      1.4\; \mbox{TeV} & \mbox{(LHC@7)} \\ 
                                      1.5\; \mbox{TeV} & \mbox{(LHC@8)}
\end{array}  \right. \quad \mbox{(95\% CL)} \,, \label{top}
\ee
respectively.

For recent bounds on the characteristic scale of these operators from measurements of dijets at the LHC, see~\cite{4f}. In these references, $\Lambda_{\rm  CI}$ in \eq{fermi} is found to be around  10 TeV in the case of light quarks. The limits (\ref{top}) coming from the top quark are less stringent but of the same order  of magnitude.

\subsubsection{Strongly interacting light Higgs models}

Strongly interacting light Higgs model (SILH) are theories in which the Higgs multiplet is assumed to belong to a new (strong) sector responsible for the EW symmetry
breaking \cite{Giudice:2007fh}. These models are broadly characterized by two parameters, a coupling constant $g_*$ and a scale $m_*$, which denotes the mass of the heavy physical states. 

The leading new physics effects are parametrized in terms of dimension-6  operators, involving the Higgs and the other SM fields, consistent with $SU(3)_C\times SU(2)_L\times U(1)_Y$ symmetry.  Among these operators, we are interested in the following one
\be\label{ctg}
c_{tG} \; \bar q_L H^c \sigma ^{\mu\nu} T_A t_R G^A_{\mu\nu} + \mbox{h.c.}\, ,
\ee
where $q^T=(t\,\,b)$ and $H^c=i\sigma^2 H^*$ is the conjugated Higgs field. The size of the coefficients of these effective operators is usually derived by naive dimensional analysis (NDA)~\cite{NDA} as described in \cite{Giudice:2007fh,Contino:2013kra}. 

Naive estimation of the coefficient $c_{tG}$ in \eq{ctg} gives $c_{tG}\sim {g_s y_t}/{m_*^2}$.
After EW symmetry breaking, the operator in \eq{ctg} reduces to the operator $\hat O_2$ of \eq{O} with the identification 
\be \label{id}
{|C_2|}/{2m_t^2}={y_t}/{m_*^2}\,.
\ee
In this way we can translate the bounds on $C_2$ into limits on the mass $m_*$. By using the results of the previous section we find 
\be
m_\star > \left\{  \begin{array}{cc} 
1.2\; \mbox{TeV} & \mbox{(Tevatron)} \\ 
0.9\; \mbox{TeV} & \mbox{(LHC@7)} \\ 
0.9\; \mbox{TeV} & \mbox{(LHC@8)}
\end{array}  \right. \quad \mbox{(95\% CL)} \,. \label{mstar}
\ee

A special phenomenological realization of SILH is represented by theories where the Higgs doublet is a composite Goldstone boson of a spontaneously-broken symmetry of the strong dynamics~\cite{compo,Agashe:2004rs}. In these models the ratio $m_*/g_*\equiv f$ is  identified with the decay constant of the Goldstone boson associated with the symmetry breaking. An important quantity of these composite models is the ratio $\xi = \upsilon^2/f^2$, which characterizes the distance between the EW and the strong dynamics scales.

How the scale $m_\star$ should properly interpreted within the composite models depends on  whether the assumption of minimal coupling is taken as a guiding principle or not. For a recent discussion and criticism about this point see~\cite{Jenkins:2013fya}. 

If we assume the strongly interacting theory to be minimally coupled, the coupling of the composite top quark to the gluon field remains the same as for the SM fermions. Accordingly, the operator in \eq{ctg} can be generated only at loop level---by means of the coupling to the heavier resonances---and the coefficient $c_{tG}$ receives a further suppression by a factor of $g_*^2/16\pi^2$. In this case, while it is possible to obtain a bound on the scale $f$---because of the identification ${|C_2|}/{2m_t^2}={y_t}/{16\pi^2f^2}$---and therefore on the parameter $\xi $, the constraints we have found are  too weak to bound the parameter space of these models. 

On the other hand, if the interaction with the gluon field of the composite top quark is assumed to be non-minimal, which is the most reasonable assumption for a composite object, then we can start out directly with the  interaction vertexes in \eq{int} and $m_*$ can be taken to coincide with the mass of the heavy composite fermion. The result holds both in the case of complete and partial compositeness. A specific model of partial compositeness in which this scenario is realized is discussed in the Appendix A. 

In composite models, both the right-handed and left-handed top quark should have a sizable degree of compositeness. The right-handed top quark $t_R$ could be completely composed, there are no experimental limits beside those discussed here. The compositeness of the left-handed top quark $t_L$ is more constrained because of its pairing with the $b$ quarks in  a  $SU(2)$ doublet and the experimental constraints from the decay $Z\rightarrow b \bar b$. However, it is  possible to show~\cite{bL} how to protect $b_L$ from corrections in such a decay thus leaving open the possibility of its being a composite state as well.

In the case of the composite Higgs models, the composite fermion masses 
must be close to the scale  $f$ \cite{Pomarol:2008bh,Panico:2012uw} in order for the Higgs boson mass to be equal to its experimental value. If we assume a non-minimal coupling scenario, following the discussion above, it is possible to identify $m_\star \simeq f$ in \eq{id}. In this case the parameter $\xi$ is accordingly constrained to be
\be
\xi < \left\{  \begin{array}{cc} 0.04 & \mbox{(Tevatron)} \\ 
                                             0.07 & \mbox{(LHC@7)} \\ 
                                             0.08 & \mbox{(LHC@8)}
\end{array}  \right. \quad \mbox{(95\% CL)} \, . \label{xi}
\ee
Notice that values of $\xi$ below 0.1 requires a high degree of cancellation between different terms in order to keep the Higgs boson mass at its experimental value and are  therefore considered unnatural and make the usefulness of the model doubtful. 

In considering the above constraints, we must bear in mind that there are many caveats depending on the connection in a specific  model between the composite states and the top quark. Most notably, in models with partial compositeness~\cite{Kaplan:1991dc,Grossman:1999ra} further suppression factors---making the above limits weaker---may come from the mixing angles between heavy composite and light states, as shown in a specific model in the Appendix A. The same model  also shows that while the limits in \eq{xi} can be slightly relaxed, they cannot be made substantially weaker.

An independent limit on the scale $f$ can be obtained by means of the operator $\hat O_1$, which, using the equations of motion, gives rise to the four-fermion operator of \eq{ff}. This operator, according to NDA and minimal coupling, has a coefficient that is proportional to  $f^{-2}$ without any loop suppression. An  example is given by the typical four-fermion operator considered in SILH theories that involves only right handed top quarks:
\be\label{o4t}
c_{4t} \; \bar t_R \gamma^\mu t_R \, \bar t_R \gamma^\mu t_R \, .
\ee

A naive estimation of the coefficient $c_{4t}$ gives $c_{4t}\sim g_s^2/f^2$. A comparison of this operator with the one in \eq{ff}, after taking into account the color factors, gives the following identification:
\be 
{|C_1|}/{3m_t^2}={1}/{f^2} \, .
\ee 
In this way, we could directly translate the constraints on the coefficient $C_1$ into limits on the scale $f$ and the parameter $\xi$ which are comparable to those in \eq{xi}. However, this is only possible if we  treat all flavors on an equal footing, an assumption that does not apply to most composite models  where the coupling to the  light quarks are explicitly taken to be different and much suppressed with respect to that to the top quark.

Finally, the limits on $m_\star$ in \eq{mstar} imply that the masses of these fermion are close or larger than 1 TeV. These states are described as custodial fermions because they prevent the mass of the Higgs boson from being too large but this is only true if their masses are less than 1 TeV~\cite{Pomarol:2008bh,Panico:2012uw}.

%%%%%%%%%%%%%%%%%%%%%%

\begin{acknowledgments}
MF is associated to SISSA. He thanks M. Serone for discussions on  composite Higgs models and NORDITA in Stockholm for the hospitality during the completion of this work. The work of AT 
was supported by the S\~ao Paulo Research Foundation (FAPESP) under grants 2011/11973-4 and 2013/02404-1. He thanks G. von Gersdorff for useful discussions and the ICTP Trieste for the hospitality during the completion of this work.  
\end{acknowledgments}

%%%%%%%%%%%%%%%%%%%%%%%%%%%%%%%%%%%%%
\appendix
\section{Partial compositeness and non-minimal coupling}

In this appendix we introduce an explicit model of partial compositeness to show how the operators in \eq{O} cannot be rotated away at the tree level by a field redefinition if we assume non-minimal coupling. In the process, we also obtain an estimate of the additional suppression generated by the mixing between composite and SM fermions.

Many realizations of composite Higgs models rely on the hypothesis of partial compositeness~\cite{Grossman:1999ra}, in which each SM state has a composite partner with equal quantum numbers under the SM symmetries. These fields are multiples of the global symmetry of the composite sector which can be taken minimally to be $SU(3)_c\times SU(2)_L\times SU(2)_R \times U(1)_X$.  We consider here a simplified case in which the composite  partners of the top quark are vector-like fermions belonging to the representations $\Psi\sim (2,2)_{2/3}$ and $\tilde T \sim (1,1)_{2/3}$ of  $SU(2)_L\times SU(2)_R\times U(1)_X$.
The multiplet $\Psi$ 
\be
\Psi = 
\left(  \begin{array}{cc}   
T & X_{5/3} \\ 
B & X_{2/3}                                     
\end{array}  \right)
\ee
contains a doublet $Q=(T\,\,B)^T$ with the same quantum numbers of the  SM left-handed doublet $q^{el}_L=(t^{el}_L\,\,b^{el}_L)^T$, while $\tilde T$ has the same quantum numbers of the SM right-handed top $t^{el}_R$. 

The composite fermion states have an explicit Dirac mass term and are assumed to mix linearly with the SM elementary fields as in the following lagrangian:
\bea 
{\cal L}_{mass}+{\cal L}_{mix}&=&-M_Q \Tr \bar \Psi \Psi - M_{\tilde T} \bar {\tilde T}\tilde T-\Delta_Q \bar Q_R  q_L^{el} - \Delta_{\tilde T} \bar {\tilde T}_L t^{el}_R  +\mbox{h.c.}\nn\\
&=&-\bar T_R(M_Q T_L+\Delta_Q t_L^{el})-\tilde T_L (M_{\tilde T}\tilde T_R+ \Delta_{\tilde T}t^{el}_R )+\mbox{h.c.} +\ldots
\eea

The mass mixing arising from ${\cal L}_{mix}$ can be diagonalized by the following field transformations:
\be \label{dL}
\left\{  \begin{array}{cc}
t_L^{el}=&\cos{\varphi_L}\, t_L + \sin{\varphi_L}\, T'_L\\
T_L=&-\sin{\varphi_L}\, t_L + \cos{\varphi_L}\, T'_L\end{array} \right.
\qquad\mbox{and}\qquad
\left\{  \begin{array}{cc}
t_R^{el}=&\cos{\varphi_R}\, t_R + \sin{\varphi_R}\, \cal T_R\\
\tilde T_R=&-\sin{\varphi_R}\, t_R + \cos{\varphi_R}\, \cal T_R
\end{array} \right.\,.
\ee
The top fields $t_L$ and $t_R$ are the massless (before EW symmetry breaking) partially-composite eigenstates, while $T'_L$ and $ \cal T_R$ are the massive composite ones, their  masses being $M_{T'}=M_Q\cos{\varphi_L}+ \Delta_Q\sin{\varphi_L}$ and $M_{\cal T}=M_{\tilde T} \cos{\varphi_R}+ \Delta_{\tilde T} \sin{\varphi_R}$. The mixing angles $\varphi_L$ and $\varphi_R$ are defined such that ${\rm tan}\varphi_L=\Delta_Q/M_Q$ and ${\rm tan}\varphi_R=\Delta_{\tilde T}/M_{\tilde T}$.

QCD gauge fields are coupled to the fermions through the covariant derivative 
$\slashed D=\gamma^\mu(\partial_\mu - ig_s T_A G^a_\mu)$ in the kinetic terms:
\bea 
{\cal L}_{kin}&=&\bar q^{el}_L i\slashed D q^{el}_L+\bar t^{el}_R i\slashed D t^{el}_R
+\bar Ti \slashed D T+\bar {\tilde T} i\slashed D \tilde T +\ldots \, .
\eea
EW interactions are not relevant for our discussion and they are explicitly set to zero. Once the rotation of the fields into the mass eigenstates is performed, using \eq{dL}, the lagrangian reads:
\bea 
{\cal L}_{kin}+{\cal L}_{mass}+{\cal L}_{mmix}&=&\bar t_L \slashed D t_L+\bar T'_L \slashed D T'_L+
\bar t_R \slashed D t_R+\bar {\cal T}_R \slashed D {\cal T}_R + \bar T_R \slashed D T_R+
\bar {\tilde T}_L \slashed D \tilde T_L\nn\\
&-&M_{T'}(\bar T_RT'_L+h.c.)-M_{\cal T}(\bar {\tilde T}_L{ \cal T_R}+h.c.)+\ldots
\eea

If the composite sector is assumed to be minimally coupled then the chromo-magnetic operator in \eq{ctg} can only be generated at loop level (for an explicit example see Appendix A of \cite{Redi:2013eaa}). The same holds for the other operator in \eq{O}.

On the other hand, if we allow the composite sector to be non-minimally coupled, then it is possible to introduce the following chromo-magnetic interaction term among the composite fermions:
\be 
{\cal L}'=\kappa \, \bar Q H^c \sigma^{\mu\nu} G_{\mu\nu} \tilde T +{\rm h.c.}\,,
\ee
where $H^c$ is the conjugated composite Higgs field. This operator is suppressed by two inverse powers of the composite fermion mass, namely the coefficient $k$ can be taken to be $\kappa \sim g_sy_t/M_Q^2$, where $y_t$ is the Yukawa coupling of the top quark. After EW symmetry breaking, by performing the rotation  \eq{dL}, we obtain
\bea
{\cal L}'&=&\sin{\varphi_L} \sin{\varphi_R} \frac{g_sy_t}{M_Q^2}\, \upsilon\,\bar t_L \sigma^{\mu\nu} G_{\mu\nu}t_R 
+{\rm h.c.}+\ldots\, \label{xx}
\eea
where the dots stand for  other terms which involve the heavy composite fields. A similar argument could be made for the other operator in \eq{O}. 

Comparing \eq{xx} with \eq{id}, we have the following identification:
\be 
m_*^2=M_Q^2/\sin{\varphi_L} \sin{\varphi_R} \,. \label{sup}
\ee
Eq.~(\ref{sup})  shows that limits  on $M_Q$ turn out to be weaker than  those on $m_*$  because of the presence of the suppression factor given by the sine of the mixing angles. However, these mixing angles cannot be too small because they enter into the definition of the top Yukawa coupling, which is generated from the following interaction: 
\be 
{\cal L}_{Yuk}=Y^*\bar Q H^c \tilde T + {\rm h.c.}  \, .
\ee
Using \eq{dL}, one finds that the top Yukawa coupling is given by
\be 
y_t=\sin{\varphi_L}Y^*\sin{\varphi_R}\,.
\ee
Perturbative control on the strength of the composite interaction requires that $Y^*\lesssim 3$. Then, in order to have $y_t\sim 1$, one needs sizeable mixing angles $\varphi_L$ and $\varphi_R$.

%%%%%%%%%%%%%%%%%%%%%%%%%%%%%%%%%%%%%%

%%%%%%%%%%%%%%%%%%%%%%%%%%%%%%%%%%%%%%%%%%%%%%

\end{document}